\documentclass[useAMS,usenatbib]{mn2e}

\usepackage{graphicx}
\usepackage{latexsym}
\usepackage{amssymb}
\usepackage{amsmath}
\usepackage{verbatim}
\usepackage[color]{showkeys}

\newcommand\eiso{\ensuremath{E_{\mathrm{iso}}}}
\newcommand\ep{\ensuremath{E_{\mathrm{p,i}}}}
\newcommand\sigext{\ensuremath{\sigma_{\mathrm{ext}}}}
\newcommand\dedm{dark energy and dark matter}

\def\lsim{\mathrel{\rlap{\lower4pt\hbox{\hskip1pt$\sim$}}
    \raise1pt\hbox{$<$}}}
\def\gsim{\mathrel{\rlap{\lower4pt\hbox{\hskip1pt$\sim$}}
    \raise1pt\hbox{$>$}}}
\def\sqr#1#2{{\vcenter{\vbox{\hrule height.#2pt
         \hbox{\vrule width.#2pt height#1pt \kern#1pt
         \vrule width.#2pt}
         \hrule height.#2pt}}}}

\def\beq{\begin{equation}}
\def\eeq{\end{equation}}
\def\beqa{\begin{eqnarray}}
\def\eeqa{\end{eqnarray}}
\begin{document}

\title[Gamma-Ray Bursts and DE - DM interaction]{Gamma-Ray Bursts and Dark Energy - Dark Matter interaction}
\author[T. Barreiro, O. Bertolami and P. Torres]{
T. Barreiro$^{1}$\footnotemark[1]\thanks{Also at Instituto de Plasmas e Fus\~ao Nuclear, Instituto Superior T\'ecnico, Lisboa. Email address: tiagobarreiro@fisica.ist.utl.pt}, 
O. Bertolami$^{2}$\footnotemark[1]\thanks{Also at Instituto de Plasmas e Fus\~ao Nuclear, Instituto Superior T\'ecnico, Lisboa. Email address: orfeu@cosmos.ist.utl.pt} and 
P. Torres$^{2}$\footnotemark[1]\thanks{Also at Centro de F\'isica Te\'orica e de Part\'iculas, Instituto Superior T\'ecnico, Lisboa. Email address: torres@cftp.ist.utl.pt}\\
$^{1}$Dept. de Matem\'atica, Univ. Lus\'ofona de Humanidades e Tecnologias, Av. Campo Grande, 376, 1749-024 Lisboa, Portugal\\
$^{2}$Departamento de F\'isica, Instituto Superior T\'ecnico, 
Av. Rovisco Pais 1, 1049-001 Lisboa, Portugal
}

\date{}
\pubyear{2010}

\maketitle

\begin{abstract}
In this work Gamma Ray Burst (GRB) data is used to place constraints on a putative coupling between dark energy and dark matter. Type Ia supernovae (SNe Ia) constraints from the Sloan Digital Sky Survey II (SDSS-II) first-year results, the cosmic microwave background radiation (CMBR) shift parameter from WMAP seven year results and the baryon acoustic oscillation (BAO) peak from the Sloan Digital Sky Survey (SDSS) are also discussed. The prospects for the field are assessed, as more GRB events become available.
\end{abstract}


\section{Introduction}

The nature of dark energy  and dark matter  remains an outstanding open problem in cosmology. In spite of the success of the $\Lambda$CDM parameterization, one must consider more complex models in order to cast some light on the substance of the dark components of the universe.
In this work one considers models with interacting \dedm\ components.
There are several useful tools to probe the phenomenology of these models, such as CMBR data \citep{wmap7}, BAO \citep{BAOprim, BAOsec}, SNe data \citep{SneSDSS} and the deviation from the virial equilibrium of galaxy clusters \citep{intmodel1,intmodel2,abdalla2009}.
It has been suggested \citep{schaefer2002,bertolamisilva2006} that GRB may be used to extend the Hubble diagram to high redshifts, greater than $z=5$. At these epochs the Universe was dominated by dark matter, from which follows that this tool is less sensitive to dark energy. However, for models where dark energy and matter are coupled \citep{amendola,intmodel1} or unified \citep{kamenshcik,GCGprimer,bento2003,GCGwmap5}, GRBs might be a particularly usefull tool \citep{bertolamisilva2006}.

In the late 1960s, the Vela array of military satellites detected flashes of radiation originating in apparently random directions in space.
The observed bursts lasted between tens of milisecond and thousands of seconds, and were composed of soft ($0.01$ to $1~MeV$) gamma rays.
Subsequently space missions such as the US Apollo program and the Soviet Venera probes confirmed the existence of the GRBs, even though its rate of occurrence was virtually unknown until the deployment of the Compton Gamma Ray Observatory, in 1991. This observatory was equipped with a sensitive gamma-ray detector, the Burst and Transient Source Explorer (BATSE) instrument which was able to detect one or two events per day.
The collected data allowed to divide GRBs into two categories: short duration bursts ({\it short} bursts) and long duration bursts ({\it long} bursts). The former usually last for less than two seconds and are dominated by high energy photons; the latter last longer than two seconds and are dominated by lower energy photons. However, this distinction is not always clear.

The physical origin of GRBs has been debated for a long time, before their exact position and a reliable estimate of their distance was lacking (see e.g. \citep{bertolami1999} and references therein). 
In 1997, several GRBs were detected by the BeppoSAX sattelite. A GRB {\it prompt} emission is followed by an {\it afterglow} emission composed by all wavelengths. Depending on its brightness, an afterglow can last from days to months after the burst itself, the {\it transient} phase.
The detection of the afterglow did manifold the information on GRBs.
Through their afterglow, GRB's X-ray, optical and radio counterparts were observed, as well as their redshifts \citep{review1}, confirming the cosmological origin of most, if not all, the GRBs.
When, in 2003, the long GRB 030329 was discovered and linked with the supernova SN2003dh \citep{GRB&SN}, it became clear that GRBs are linked with the release of gravitational energy during the collapse of stellar mass objects. GRBs are most likely collimated, considering they reach integrated
luminosities up to $L \sim 10^{53}~erg~s^{-1}$, making it hard to associate them with an astrophysical object otherwise.
This high energy release creates an outflow that expands relativistically. Two forms of shocks are ensued by the burst, the forward shock and the reverse one, which one separated by a contact discontinuity \citep{GRBprimer}. If the ejected plasma is too strongly magnetized, only the forward shock is formed.
One suggested possibility is that the prompt emission is generated in a baryon dominated ejecta through internal shocks, while the forward and reverse shocks yield the long lasting broadband emission, the afterglow \citep{GRBprimer}.
Actually, the full understanding of the prompt emission mechanism, a basic GRB property, is still lacking. One possibility is that the prompt emission consists of synchrotron radiation \citep{GRBprimer}, by the relativistic charged particles moving on the magnetized ejected plasma.
Currently, the GLAST/FERMI mission, in operation, is continuously increasing the available GRB data and making it worth, as will be discussed and pursued in this work, considering future prospects for the subject. For an overview of most recent missions see e.g. \citep{mcbreen} and references therein.

GRBs can be used as distance indicators \citep{ghirlanda2004,bertolamisilva2006,liang&zhang,amati2008}. Its main attractiveness is that the redshift range extends much higher than that of SNe Ia.
The main observables that can be measured when studying GRBs are its spherical equivalent energy, its peak isotropic luminosity, the peak energy of its spectrum, the photon fluence, the energy fluence, the pulse duration and the redshift of its host galaxy.
Several empirical correlations among these variables can be established. However, there are still large uncertainties in their calibration. Furthermore, there is still no satisfying physical mechanism accounting for them, so that assuming that they hold true can introduce systematic uncertainties in our distance indicator. 
From the existing correlations, the very discussed Ghirlanda relation uses the peak energy of the spectrum, \ep and the collimation corrected energy, $E_{\gamma}$ \citep{ghirlanda2004}. 
On the other hand, the Liang and Zang relation correlates the isotropic equivalent energy, \eiso~with \ep~and the jet break time of the afterglow of the burst \citep{liang&zhang}.
Finally, the Amati relation, correlates the isotropic energy, \eiso, with \ep~\citep{amati2008}.
This relation is particularly interesting since the \ep~- \eiso~correlation requires only two parameters that can be directly inferred from the obervations. This correlation further emphasizes the relevance of the GRB data.
Notice that the aforementioned synchrotron process reproduces the Amati correlation, a quite interesting feature. 

In this work, GRB data and the Amati relation, in particular, are used to probe a generic dark energy - dark matter interacting model.
In section \ref{DEDMint}, the interacting model is presented. The Amati \ep~- \eiso~correlation is introduced and discussed in section \ref{dataanalysis}. The set of real GRB data is then extended to a mock sample of 500 GRBs using a method detailed in subsection \ref{genmock}. In section \ref{method}, one discusses the constrains obtained from SNe data in subsection \ref{SNs}, BAO in subsection \ref{BAOs} and CMBR shift parameter in subsection \ref{CMBs}. 
In section \ref{results} one presents the obtained results. In section \ref{conclusion}, conclusions are presented.

\section{Dark energy and dark matter interaction}\label{DEDMint}

The cosmological model consists of homogeneous matter (dark matter and baryons) and dark energy, where the dark matter and energy are interacting and have equations of state, $p_{DM} = 0$ and $p_{DE}=w\rho_{DE}$, respectively.
The coupled energy densities with a coupling $\zeta$ evolve as follows \citep{intmodel1}:
\beqa
\dot{\rho}_{DM} + 3 H \rho_{DM} &=& \zeta H \rho_{DM}~,
\label{bianchi1}\\
\dot{\rho}_{DE} + 3 H \rho_{DE} (1 + w) &=& -\zeta H \rho_{DM}~.
\label{bianchi2}
\eeqa

The analysis assumes for the ratio of the dark components that \citep{intmodel1}
\beq
\frac{\rho_{DE}}{\rho_{DM}} = \frac{\Omega_{DE_0}}{\Omega_{DM_0}} a^\eta = \frac{\Omega_{DE_0}}{\Omega_{DM_0}} (1+z)^{-\eta}~,
\label{DMDEratio}
\eeq
\noindent for a constant $\eta$, where, $a$, is the scale factor, assumed that at present $a_0=1$, and $z$ is the redshift.

Inserting the time derivative of Eq. (\ref{DMDEratio}) into Eqs. (\ref{bianchi1}) and (\ref{bianchi2}), one obtains for the coupling
\beq
\zeta = \frac{\zeta_0}{\Omega_{DE_0} + \Omega_{DM_0} (1+z)^{\eta}}~,
\label{eta0}
\eeq
where $\zeta_0=-(\eta + 3 w)\Omega_{DE_0}$. Note that when $\eta = -3 w$, there is no interaction between \dedm, since $\zeta=0$.

The solutions for Eqs. (\ref{bianchi1}) and (\ref{bianchi2}) can be written as
\beq
\rho_{DM} = (1+z)^{3} \rho_{DM_0} \Bigl[\frac{\Omega_{DE_0}(1+z)^{-\eta} + \Omega_{DM_0}}{\Omega_{DE_0} + \Omega_{DM_0}}\Bigr]^{\frac{-\eta-3w}{\eta}}~.
\label{solDM}
\eeq

\beq
\rho_{DE} = (1+z)^{-\eta+3} \rho_{DE_0}  \Bigl[\frac{\Omega_{DE_0}(1+z)^{-\eta} + \Omega_{DM_0}}{\Omega_{DE_0} + \Omega_{DM_0}}\Bigr]^{\frac{-\eta-3w}{\eta}}~.
\label{solDE}
\eeq

Inserting Eqs. (\ref{DMDEratio}), (\ref{solDM}) and (\ref{solDE}) into the Friedmann equation for a flat universe, $H^2=H_0^2 (\rho_{DM}+\rho_{DE}+\rho_b)/\rho_0$, where $H_0$ and $\rho_0$ are the 
Hubble constant and the total energy density at present, one obtains
\beq
E^2(z) = (1+z)^3\left[\frac{[\Omega_{DM_0}+\Omega_{DE_0}(1+z)^{-\eta}]^{\frac{-3w}{\eta}}}{(\Omega_{DM_0}+\Omega_{DE_0})^{\frac{-\eta-3w}{\eta}}}+\Omega_{b_0}\right]~,
\eeq
\noindent where $\Omega_{b_0}$ and $\Omega_{DM_0}$ are the baryon and DM energy densities at present and $E(z) = H(z)/H_0$.

It is interesting to point out that the Generalized Chaplygin Gas model (GCG) \citep{kamenshcik,GCGprimer}, an unified model of dark energy and dark matter, can be seen as a particular case of this interacting model for $\eta = 3(1+\alpha)$ and $w = -1$ \citep{bento2004}, $\alpha$ being the GCG equation of state parameter, $p = -\frac{A}{\rho^\alpha}$, where $A$ is a positive constant.

\section{Gamma Ray Burts}\label{dataanalysis}

In this work one considers the Amati \ep~- \eiso~correlation (for a discussion of the use of other correlations see for instance \citep{bertolamisilva2006,liang&zhang}).
This correlation can be used to place constrains on the Hubble diagram. 
The sample provided in \citep{amati2008,amati2009} that includes the observations of $95$ GRB with measurements for \ep, \eiso~and redshift, is adopted.

The value of \ep~is an observable quantity, independent of a cosmological model. On the other hand, \eiso~is computed for each GRB from its spectral parameters, fluence and redshift using a specific cosmological model (The \eiso~presented in \citep{amati2008,amati2009} is computed in the context of the $\Lambda$CDM scenario with $h=0.7$, $\Omega_{M_0}=\Omega_{DM_0}+\Omega_{b_0}=0.3$ and $\Omega_{DE_0}=0.7$). The Amati correlation assumes a power law relationship between \ep~and \eiso. A cosmological model can then be tested comparing the \eiso~computed from the observations with  a theoretical \eiso~obtained from \ep. In practice, however, since the \ep~-~\eiso~relationship is not calibrated, one has to simultaneously fit for the power law parameters.

Furthermore, following \citep{amati2006}, the scatter of the \ep-\eiso~relation that cannot be explained by statistical fluctuations alone must be taken into account. As in \citep{agostini2005,amati2006}, this is done by introducing a third parameter, an extrinsic variance, \sigext, in the fitting of the data. Using a power law \ep-\eiso~relation,
\begin{equation}
\log \ep = m \log \eiso + q \;,
\end{equation}
one aims to minimize the likelihood function, where
\begin{equation}
\chi^2_\mathrm{GRB} = - \sum_{\substack{\mathrm{GRB} \\ \mathrm{objects}}} \left[
\log \left( 2 \pi \sigma^2 \right)
+ \frac{(\log E_{\mathrm{p}, i} - m \log E_{\mathrm{iso}} - q)^2}{\sigma^2} 
\right]\; ,
\end{equation}
and the total variance is $\sigma^2 = \sigext^2 + \sigma_{\mathrm{p}}^2 + m^2 \sigma_{\mathrm{iso}}^2$, where $\sigma_{\mathrm{p}}$ and $\sigma_{\mathrm{iso}}$ are the observational variances on $\log \ep$~and $\log \eiso$, respectively. The fit is performed for the three parameters $m$, $q$ and \sigext.

This minimization is then carried out for different values of the cosmological parameters, resulting in a profile of the likelihood function.


\subsection{Generating a GRB mock sample}\label{genmock}

Given that the GRB data is currently rather limited, the analysis is extended to include a mock sample of GRBs in order to test the efficiency of the Amati relation on constraining \dedm\ interacting models. The goal is to check the effect of a larger number of GRBs, but also the effect of higher redshift GRBs. 
\textbf{From 2009 onwards, most of the useful events to fit the Amati relation came from the Swift and Fermi experiments. One can expect from these experiments approximately 10 useful GRB events per year. The future launch of the EXIST mission, scheduled for 2017, will considerably improve this rate, hopefully simultaneously reducing the measurement errors. This paper settles on a best case scenario of 500 GRBs events but conservatively keeping the error bars at the present level.} 

Following \citep{amati2008}, a distribution mimicking the observed GRB redshift distribution in the range $0< z<6$ is used. A chosen percentage of these events was then replaced with redshifts uniformly distributed in the range $6 < z < 10$. This allows for a tuning of the number of high redshift GRBs in the mock sample that is used in the fits.
Notice that different choices on the shape of the high redshift distribution of GRBs is approximately equivalent to a change on the redshift cutoff value and its percentage. Ultimately, only a significant number of high redshift GRBs can yield a sizable restriction on $\zeta_0$.
 The details about the actual low redshift distribution used to generate the mock data are discussed in the Appendix.
 
Once a redshift distribution is obtained, lognormal distributed values of \ep~are attributed to each data point and are associated with a power law related \eiso. In this paper a value of $5.86$ for $\log (\ep/1 \mathrm{keV})$ is used for the mean and a value of $1$ is used for the variance. An extrinsic variance (using $\sigma_\mathrm{ext} = 0.41$) and Gaussian errors for \ep~and \eiso~($20\%$ for both) are then included, mimicking the current observational situation. It is verified that the results are not particularly sensitive to these choices. 
Several mocks with a varying number of low and high redshift GRBs where generated and studied. The presented results consist of a typical mock sample with $500$ events generated and $10\%$ of high redshift GRBs, in a $\Lambda$CDM universe.

\section{Cosmological data}\label{method}

In order to gauge to which extent GRB data can constrain cosmological parameters, one confronts it with other well known cosmologically relevant observational tests such as SNe, BAO and the CMBR shift parameter.

\subsection{Supernovae}\label{SNs}

The SNe sample from the Sloan Digital Sky Survey II (SDSS-II) first-year results \citep{SneSDSS} is used, consisting of $288$ SNe Ia with redshifts up to $1.55$.
SNe are used as distance indicators by comparing the theoretical distance modulus, $\mu_\mathrm{th}$, obtained from the measured redshift in 
a given model of cosmological evolution, and with the inferred distance modulus, $\mu_\mathrm{obs}$, computed from  fits to the SNe light curves (using the data of the \textsc{mlsc2k2}  fits found in \citep{SneSDSS}).

Specifically, the theoretical distance modulus is given by 
\beq
\mu_\mathrm{th} = 5 \log{D_L} + 5 \log(\frac{c}{H_0}) + 25\;,
\eeq
with $D_L$ being the scaled ($H_0$ independent) luminosity distance in Mpc,
\beq
D_L(z) =  (1+z) \int_0^z \frac{dz'}{E(z')}~.
\label{dl}
\eeq

For each cosmological parameter choice, the used likelihood is given by
\beq
\chi^2_\mathrm{SN} = \sum_{\substack{\mathrm{SN} \\ \mathrm{objects}}} \frac{(\mu_\mathrm{obs} - \mu_\mathrm{th})^2}{\sigma_{\mu}^2}
\eeq
where $\sigma^2_{\mu} = \sigma^2_\mathrm{fit} + \sigma^2_\mathrm{disp} + \sigma^2_\mathrm{z}$ is the measurement variance for $\mu_\mathrm{obs}$, including the error from the fit, an intrinsic dispersion error of $\sigma_\mathrm{disp} = 0.16$, and a redshift error to account for the host galaxy peculiar movement and spectroscopic measurement. Only the SNe results are marginalized over $H_0$ with a flat prior; all the other constraints assume a constant $h=0.7$. 

\subsection{Baryon Acoustic Oscillations}\label{BAOs}

One also considers the constraints from the effect of the baryon acoustic peak of the large scale correlation function at
$100 h^{-1}$ Mpc separation detected by the SDSS  Luminous Red Galaxy sample \citep{BAOprim, BAOsec}. The peak position is related to the quantity

\beq
\mathrm{A} = \sqrt{\Omega_{M_0}} E(z_1)^{-1/3} \left[  \frac{1}{z_1}  \int_{0}^{z_1} \frac{dz}{E(z)}   \right]^{2/3}~,
\eeq
measured to be $A_0 = 0.493$, with an error of $\sigma_A = 0.017$. The used likelihood is given by

\beq
\chi^{2}_{\textrm{BAO}} = \left( \frac{A_0 - A}{\sigma_A} \right)^2~.
\eeq

\subsection{Cosmic Microwave Background Radiation shift parameter}\label{CMBs}

Here, the constraints from the CMBR WMAP7 observations \citep{wmap7} are considered. The shift parameter \citep{bond1997} can be used as a distance prior to constrain a given dark energy model. The shift parameter is given by the equation
\beq
R_\mathrm{th} =  \frac{D_L(z_{\star})}{(1 + z_\star)} \sqrt{\Omega_{M_0}}~,
\eeq
with $D_L$ being the luminosity distance defined in Eq. (\ref{dl}), and $z_\star$ the redshift at decoupling. The standard fitting formula for $z_\star$ is used \citep{Hu1996}. This theoretical prediction is then constrained through the fitted WMAP7 observation, $R_\mathrm{obs} = 1.725$ with error $\sigma_R = 0.018$, through
\beq
\chi^{2}_{\textrm{CMB}} = \left(  \frac{R_\mathrm{th} - R_\mathrm{obs}}{\sigma_R} \right)^2~.
\eeq
 

\section{Results and Constraints}\label{results}

\begin{figure}
\begin{center}
\includegraphics[width=9cm]{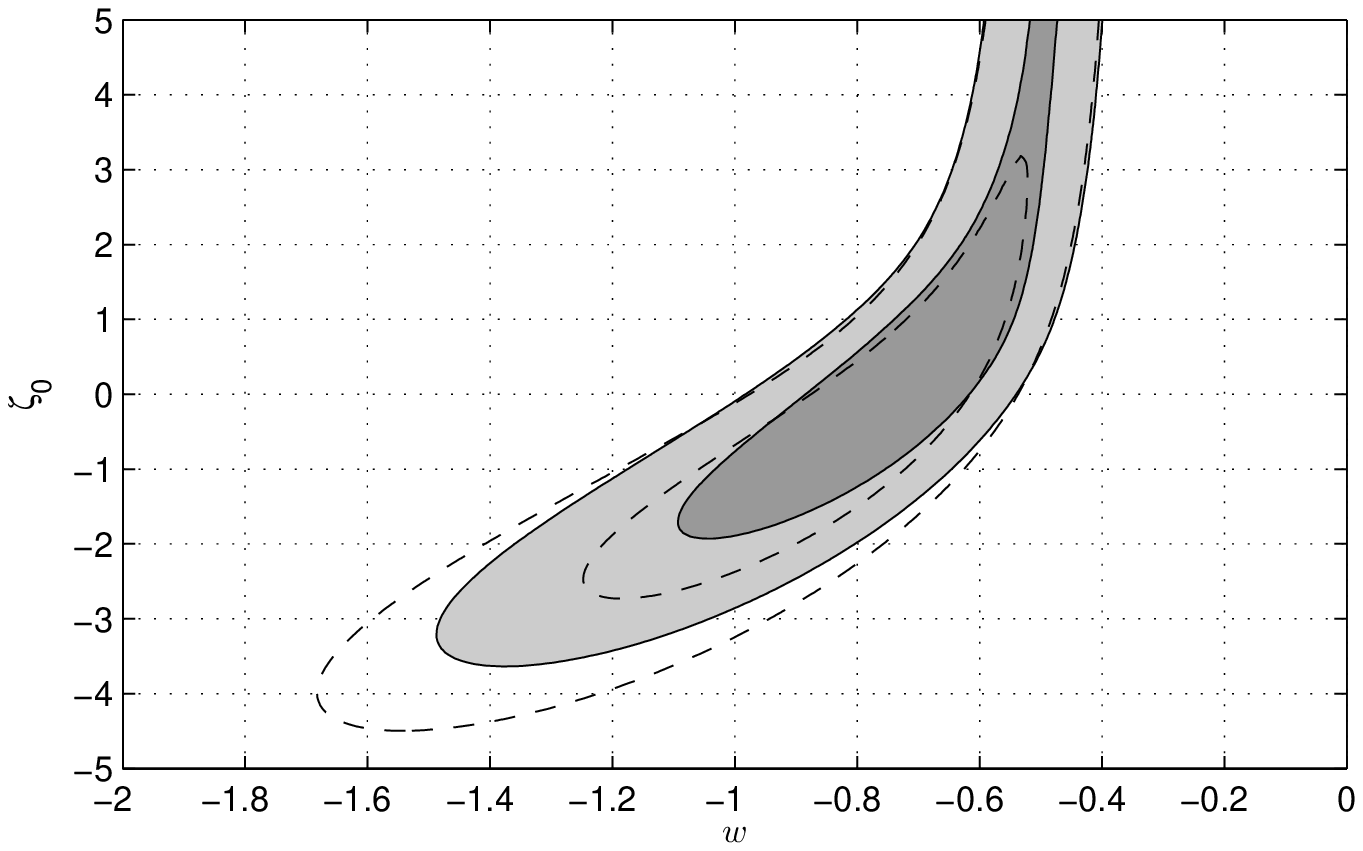}
\end{center}
\caption[fig:GRBdata]{Constraints on the values of $w$ and $\zeta_0$ obtained from the SNe data (dashed line) and its combination with the 95 GRB data (full line). 
The $68\%$ and $90\%$ confidence levels are depicted.
One fixes: $\Omega_{M_0} = 0.3$ and $h=0.7$.
}
\label{fig:GRBdata}
\end{figure}

One starts with the constraints obtained from the real observed 95 GRBs data \citep{amati2008,amati2009}, combined with the SNe SDSS-II data \citep{SneSDSS}. For this, $\Omega_{M_0}=\Omega_{DM_0}+\Omega_{b_0}$ is fixed at $\Omega_{M_0} = 0.3$ with $\Omega_{b_0}=0.0445$ and the Hubble parameter held at $h = 0.7$. The result can be seen in Fig.~\ref{fig:GRBdata}. 
At $68\%$ confidence level, the SNe data alone provides the limits $\zeta_0\in[-1.93, 1.02]$ and $w\in[-1.07, -0.62]$. 
These results are, by themselves, highly degenerate in $\zeta_0$. Including the GRB real data, one improves slightly the constraints to $\zeta_0\in[-1.15, 1.66]$ and $w \in[-0.94, -0.58]$.
As expected, the status of the GRB data at present does not allow for a significant improvement over the SNe constraints on these two parameters (see Fig. \ref{fig:GRBdata}). 
This SNe degeneracy is usually lifted combining the SNe data with the CMBR WMAP7 shift parameter observations yielding the bounds $\zeta_0\in[-0.01, 0.13]$ and $w\in[-0.83, -0.65]$ at 68\% confidence level (not shown).
Even though the present GRB data cannot compete with these CMBR constraints, with additional data they can provide an important independent method of lifting the SNe degeneracy in $\zeta_0$.
It must be stressed that these results are obtained for $\Omega_{M_0}$ fixed at $0.3$.

\begin{figure}
\begin{center}
\includegraphics[width=9cm]{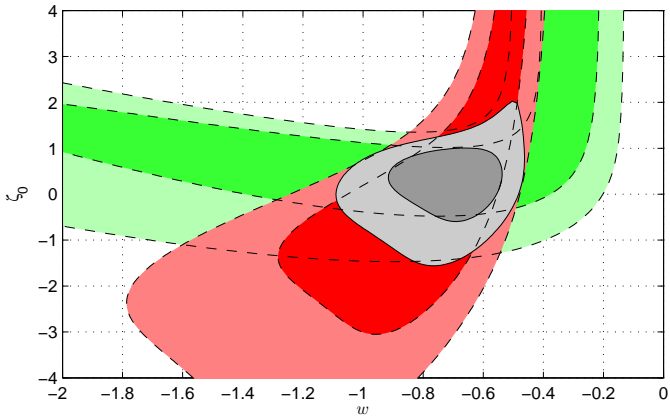}
\end{center}
\caption[fig:SNGRB]{Constraints on the values of $w$ and $\zeta_0$ obtained from SNe and a mock of 500 GRB population. 
The wider contour in $w$ corresponds to the GRBs (in green, for online readers) and the wider contour in $\zeta_0$ to the SNe (in red, for online readers). Both $68\%$ and $95\%$ confidence levels are shown. The shaded area refers to the combined confidence region.
A marginalization over $\Omega_{M_0} \in [0.2, 0.4]$ has been considered.}
\label{fig:SNGRB}
\end{figure}

To illustrate this point, a mock population is chosen in order to show what can be accomplished from a large population of GRBs, despite the current level of measurement and theoretical uncertainties. A $\Lambda$CDM universe has been used to generate a mock sample (see section \ref{genmock}). This time, the results are marginalized over $\Omega_{M_0}$ with a flat prior in the interval $0.2\leq \Omega_{M_0} \leq 0.4$.
In Fig.~\ref{fig:SNGRB} the results of the SNe and GRB constraints are shown.
The constraints obtained are $\zeta_0\in[-0.21, 0.82]$ and $w \in[-0.84, -0.58]$, at $68\%$ confidence level. 
From the SNe data alone, only a lower bound can be derived for $\zeta_0$ in the considered parameter range.

\begin{figure}
\begin{center}
\includegraphics[width=9cm]{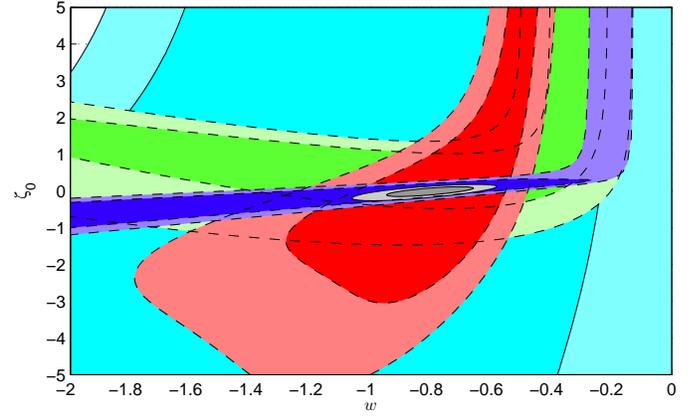}
\end{center}
\caption[fig:SNBAOGRBCMB]{The combined constraints on the values of $w$ and $\zeta_0$ obtained from the SNe (red), BAO (light blue), CMB (blue) and a mock 500 GRB population (green). The GRB and SNe contours are the same as in Fig. \ref{fig:SNGRB}. The CMB contour is the narrower one (in blue) and the BAO contour is in background (light blue). The dark patch shows the combined region. The $68\%$ and $95\%$ confidence levels are shown. 
A marginalization over $\Omega_{M_0} \in [0.2, 0.4]$ has been carried out.
}
\label{fig:SNBAOGRBCMB}
\end{figure}

Once again, this SNe degeneracy in $\zeta_0$ can be lifted by combining the SNe and CMBR constraints, yielding the bounds
$\zeta_0\in[-0.29, 0.18]$ and $w \in[-1.08, -0.65]$ at $68\%$ confidence level. 
Fig.~\ref{fig:SNBAOGRBCMB} combines all the available constraints, namely GRB, SNe, BAO and CMBR.
The CMBR data clearly provides the tighter constraints on all the parameters. 
It is, however, highly degenerate in $w$ and, hence, the SNe constraints on $w$ are required to yield significant bounds.
Notwithstanding, the GRB data has a similar profile to the CMBR bounds, and they can provide a significant bound in $\zeta_0$. Thus, GRB data, in combination with SNe, provides an independent and compatible constraint on $\zeta_0$ and $w$.
On the other hand, the BAO result clearly does not yield a strong constraint on the value of either $w$ or $\zeta_0$ (as opposed to the SNe data), and does not significantly improve the results obtained from GRBs or CMBR.
The overall combined results give the bounds
$\zeta_0\in[-0.10, 0.08]$ and $w \in[-0.89, -0.70]$ also at $68\%$ confidence level.

The GRB constraint on $\zeta_0$ comes mostly from its high redshift valued data, whereas the low redshift valued SNe data presents a degeneracy in $\zeta_0$.
Notice however that for the GRB and CMBR data, one encounters a degeneracy in $\zeta_0$ in the form of a bend occurring at $\eta<0$ ({\it i.e.} $\zeta_0>-3 w~\Omega_{DE_0}$). This occurs as for negative $\eta$ and for high $z$, the 
evolution is dominated by the dark energy density, rendering the luminosity distance virtually independent of $\eta$. 

\section{Discussion and Conclusions}\label{conclusion}

In this work the use of GRBs as cosmological tools is considered.
It has been shown \citep{bertolamisilva2006,amati2008} that GRBs have a great potential to measure the value of $\Omega_{DM}$ independently from the CMBR constraints. 
In the current work, it is shown that the present GRB data already gives a better constraint on the \dedm\ coupling parameter $\zeta_0$ than the BAO results. 
The SNe results provide good bounds on $w$, but are more degenerate in $\zeta_0$. 
Despite of that, the experimental GRB sample is still too small to provide significant constraints on $\zeta_0$.
The combined result for SNe and real GRB data further illustrates this, yielding $\zeta_0\in[-1.15, 1.66]$, with a width of $\Delta \zeta_0=2.81$, as opposed to the combined SNe and CMBR data limit $\zeta_0\in[-0.01, 0.13]$ ($\Delta\zeta_0=0.14$), both at 68\% confidence level. These results are obtained without marginalization, for fixed $\Omega_{DM_0}=0.3$.

A mock population of GRBs was then generated showing that as the number of available events increases, GRB data becomes a more and more valuable tool in constraining the parameter $\zeta_0$.
The GRBs complement the SNe constraint in a similar way the CMBR does. 
For a fixed $\Omega_{M_0}=0.3$, the SNe and mock GRB data yields $\Delta\zeta_0=0.82$.

For a deeper insight on the future cosmological implications of the GRB observations, the analysis was extended to a marginalization over $\Omega_{M_0}\in[0.2, 0.4]$.
The combined SNe and CMBR results are $\Delta\zeta_0=0.47$ (68\% CL), while for SNe and the mock GRB yields $\Delta\zeta_0=1.03$. 
Granting that the obtained bounds are not as accurate as the CMBR ones, they still provide a valuable independent measurement of these cosmological parameters.

With the same marginalization in $\Omega_{M_0}$, the combined result for SNe, CMBR and BAO is $\zeta_0\in[-0.27, 0.13]$ (95\% CL). 
The updated observations improve the previous result $\zeta_0\in[-0.4, 0.1]$ \citep{Guo}.
Note, however, that tighter priors are used in this work and that this model is slightly different, with the inclusion of non-interacting baryonic matter.

It is interesting to compare the present results with the ones arising from estimates of the departure from the virial equilibrium of the Abell cluster A586.
The bounds for $\eta$ from the Abell Cluster A586 yield $\eta\in[3.65, 4.00]$ ($\Delta\eta=0.35$), with $w=-1$ and $z=0.1708$ and $\Omega_{M_0}=0.28$, \citep{intmodel1}. 
Using SNe and mock GRB data from the present work and fixing $\Omega_{M_0}=0.28$, one encounters $\eta\in[0.93, 2.48]$ ($\Delta\eta=1.55$) at 68\% confidence level. 
If one considers the particular case of the GCG, the Abell Cluster A586 limits the $\alpha$ parameter to $\alpha\in[0.21, 0.33]$ (68\% CL). 
This compares with the combined SNe and mock GRB results, $\alpha\in[0.25, 0.83]$ (68\% CL). 
One should bear in mind that the GRB results were obtained from a mock population and are, thus, only indicative of what can be expected when the number of observed GRBs increases. 
The limits for $\alpha$ with the current observational SNe and CMBR data yield $\alpha\in[0.03, 0.06]$.

In this work, the Amati correlation \citep{amati2008} has been used, given that it requires only two parameters whose determination can be inferred and increasing the number of useful GRB events available.
Progress in the calibration and on the theoretical framework of this calibration would be invaluable, reducing the error margins in the GRB data and considerably improving the constraints on the parameters.

\section *{acknowledgments}
The work of P.T. was supported by Funda\c c\~ao para a Ci\^encia e Tecnologia (FCT, Portugal) under the grant SFRH/BD/25592/2005.

The authors would like to thank Sergio Colafrancesco, Giulia Stratta and Craig Markwardt for discussions about GRB data in the embrionary phase of this work.

\appendix

\section{GRB mock data}

For the mock sample generation the following distribution is considered \citep{porciani}
\beq
dN (P_1\leq P\leq P_2) = \frac{dV(z)}{dz}\frac{R_{GRB}(z)}{1+z} \int_{P_1,z}^{P_2,z} dL' \psi(L') \epsilon(P)~,
\label{mockdistro}
\eeq
\noindent where $dV/dz$ is the comoving volume element, $R_{GRB}$ is the comoving GRB rate density and $\epsilon(P)$ is the detector efficiency as a function of photon flux. The quantity $\psi(L)$ is the normalized GRB luminosity function and $L$ is a "isotropic equivalent" burst luminosity $L = \int_{30~keV}^{2000~keV} E~S(E)~dE$, for the energy $E$ and $S(E)$ is the rest-frame photon luminosity of the source.

The generation process is performed as follows: using a Monte-Carlo generator, a sample of the desired number of GRBs with $z$ in the range $0<z<6$ is generated, using Eq. (\ref{mockdistro}). Then, a desired percentage is replaced 
randomly by events in the range $6<z<10$, using a flat distribution. The \eiso~is randomly attributed according to a Gaussian distribution and \ep~is then calculated using a power law with the parameters calculated using a fit of the real GRB sample. 
The errors are added afterwards assuming a Gaussian distribution.


\clearpage

\end{document}